\documentclass[a4paper,12pt]{article}
\usepackage{amssymb}
\usepackage{graphicx}
\usepackage{geometry}
\usepackage{epstopdf}
\usepackage{amsmath}
\usepackage{amsfonts}
\usepackage{mathrsfs}
\usepackage{fancyhdr}
\usepackage{indentfirst}
\usepackage{amsthm}
\usepackage{abstract}
\usepackage{caption}

\title{\vspace{-1cm} A Gene Prediction Method Based on Statistics and Signal Processing}
\author{ Beilin Jia\footnote{Corresponding Author} \footnote{Shandong Univeristy, China} , Wenli Shi$^{\dag}$ , Feng Zhang$^{\dag}$}
\date{}

\begin{document}




\maketitle{\vspace{-1cm}}




\begin{abstract}
Bioinformatics, as an emerging and rapidly developing interdisciplinary, has become a promising and popular research field in 21st century. Extracting and explaining useful biological information from huge amount of genetic data is an urgent issue in post-genome era. In eukaryotic DNA sequences, gene consists of exons and introns. To predict the location of exons which carry most genetic information accurately has become one of the most essential issues in bioinformatics. Here, we have used biological characteristics of introns to find the candidate initial and final exon sections. Then we select candidate exon sections by using Support Vector Machine (SVM). Next, we predict exon sections accurately based on Discrete Fourier Transform (DFT) and using three-base periodicity of DNA sequence signals. This paper provides a gene prediction method based on statistics and signal processing and also, the improvement and prospect for this method in the future are discussed. \\

\textbf{keyword:}
Protein-coding Regions,\; Support Vector Machine,\; Discrete Fourier Transform,\; Three-base Periodicity
\end{abstract}





\section{Introduction}
\label{S:1}

Since the United States started Human Genome Project in 1990s, the sequencing of human genome and model organism genome developed rapidly. At June 26th, 2000, the work draft of human genome has been plotted. Up to now, in GenBank database, the number of bases is more than seven billion.[1] As genetic information in human genome has been interpreted gradually, we can know more about the relation between genetic information and metabolism, development, differentiation and evolution. The identification of protein-coding regions is one of the most fundamental applications in bioinformatics.\\

At present, there are dozens of methods to predict protein-coding regions. These methods can be classified into two classes, one based on sequence similarity searches and the other based on gene structure searches. Gene prediction based on the similarity of sequence utilizes similar mRNA or protein sequence, searching corresponding fragments in DNA sequence. Then this method attempts to combine similarity analysis into gene prediction. But this method depends too much on sequence homology of organisms and is limited by existing database. For to-be-testing sequences which cannot find homologous sequences, this method can be hardly realized. Gene prediction based on gene structure can be divided into two classes. One is based on statistical characters that protein-coding gene has. The other one is related to signals. These signals consist of special sequence, implying gene located around them. Prediction methods based on statistical characters is to study some statistical characters occurred in protein-coding gene. To improve accuracy of the model, they usually regard known DNA sequence as training dataset to determine model parameter. But when we do not know much about genetic information, the accuracy of identification will decrease obviously. Prediction methods based on signals find coding sequence using signal processing. Selecting appropriate DNA numerical value mapping, they use three-base periodicity and draw SNR (Signal to Noise Ratio) curve. Then they choose appropriate threshold value to identify exons. But there are still some problems in threshold value selection and windowing.\\

Therefore, combining prediction methods based on statistical characters and signals, we put forward a gene prediction method to identify protein-coding gene and locate these coding regions based on statistics and signal processing. This method enhances predicting accuracy by avoiding drawbacks of a single method and helps to interpret genetic characteristics better.

\section{Biological Background}

Gene, the basic unit of heredity in an organism, is a piece of DNA that carries genetic information. Non-gene do not code protein and has no direct relation with biology characters.[1] Discovering gene in prokaryotic genomes is less difficult, due to the higher gene density of prokaryotes and the absence of introns in their protein coding regions. Typical prokaryotic gene is illustrated by the following figure 1 .

\begin{figure}[!h]
\centering
\includegraphics[width=1\linewidth]{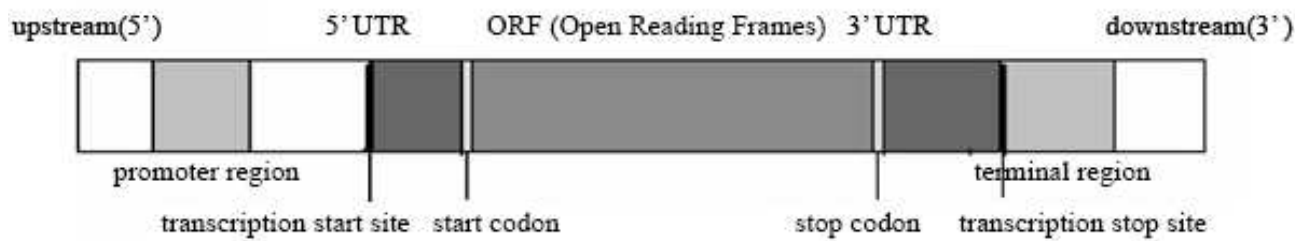}
\caption{prokaryotic gene}\label{shengwuxue1}
\end{figure}

Complete gene structure starts from promoter region and stops at terminal region. Transcription start site determines the start position of gene transcription, and transcription stops at terminal region. The content of transcription consists of 5\textquoteright  UTR (Untranslated region), ORF and 3\textquoteright  UTR. In prokaryotes, the accurate start and stop sites for translating gene are determined by start codon and stop codon and ORF, a successive coding sequence from start codon to stop codon, will be translated. But in eucaryotes, gene structure is discrete and more complicated. A typical eukaryotic gene structure is illustrated by the following figure 2 .

\begin{figure}[!h]
\centering
\includegraphics[width=1\linewidth]{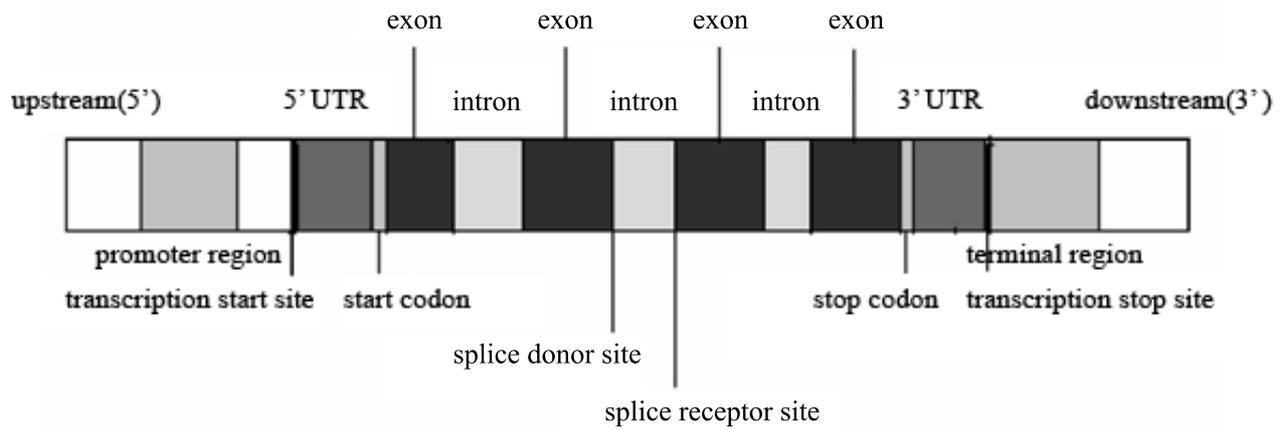}
\caption{eukaryotic gene}\label{shengwuxue2}
\end{figure}

The coding region of eukaryotic gene is not successive. Introns, the non-coding regions, divide coding regions into several pieces, the exons. So in eucaryotes, there is no ORF with definite length unlike prokaryotes. In eucaryotes, the proportion of coding sequences in the whole gene is relative small but those non-coding sequences have a large proportion, which is why eukaryotic gene structure is more complicated. Taking human genome as an example, the proportion of protein-coding genes in whole sequences is only $3\%$ to $5\%$. So predicting protein-coding regions in a new DNA is an important issue

\section{Method}
\subsection{Algorithm Process and Overview}

\begin{figure}[!h]
\centering
\includegraphics[width=0.5\linewidth]{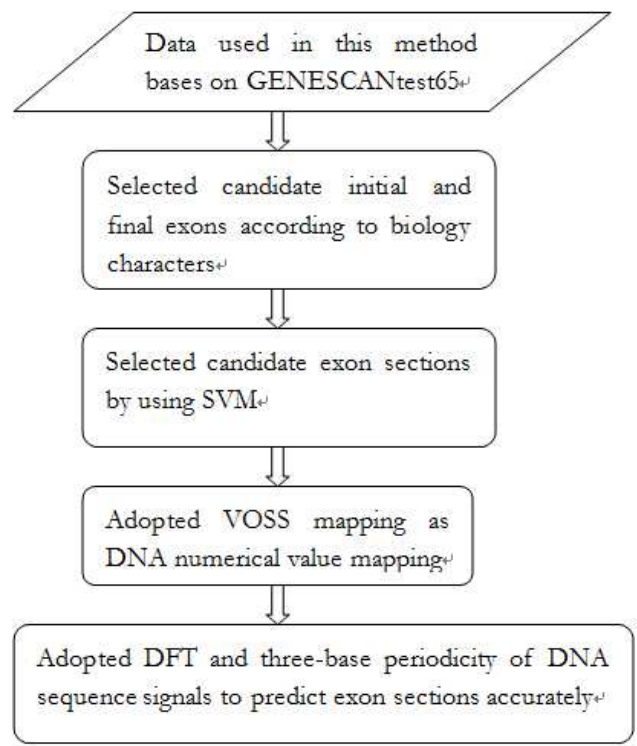}
\caption{algorithm flowchart}\label{liuchengtu}
\end{figure}

In the first step, through programming, we obtain data that includes exon sequences, length of exons, the number of each kind of bases, A, C, G, T, and of gene sequence in GENBANK. Also, we establish database of exons to make our analysis more convenience.\\

In the following steps, we design software to calculate the number of each kind of bases and seek certain sequence. Our software is efficient in extensibility, accomplishing VOSS mapping and DFT and succeeding exon prediction based on signal processing visually.

\subsection{Selection According to Biology Characters}
Introns have GT-AG character, that is, intron sequence normally begins with GT and ends up with AG. So we select sequences that have a start with GT or AG. Firstly, we draw following figures (figure 4 and figure 5) to consider the length distribution of $36832$ exons on chromosome $1$ of human.

\begin{figure}[!h]
\centering
\includegraphics[width=1\linewidth]{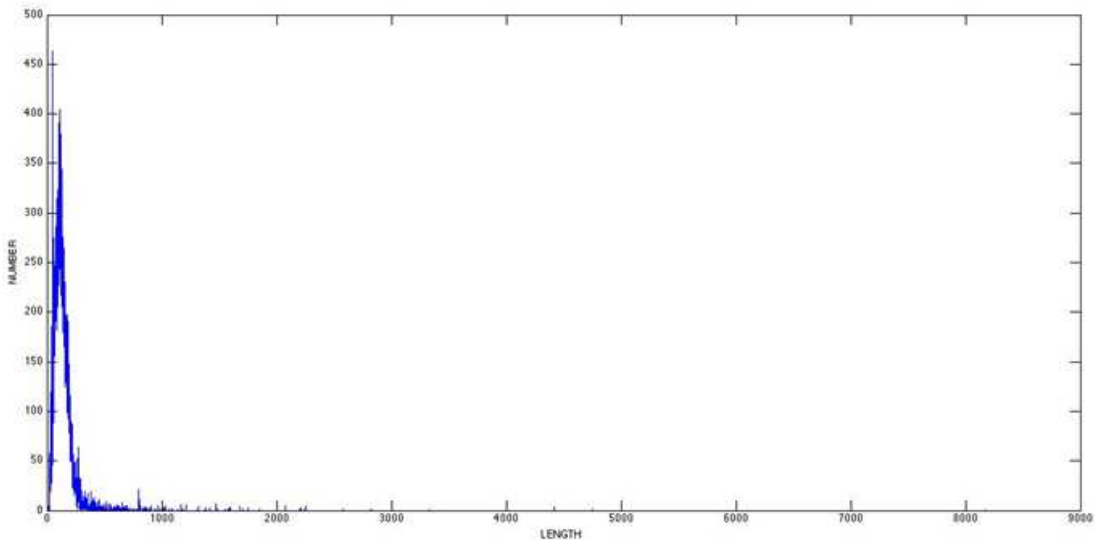}
\caption{length distribution of exons}\label{fangfa1}
\end{figure}

\begin{figure}[!h]
\centering
\includegraphics[width=1\linewidth]{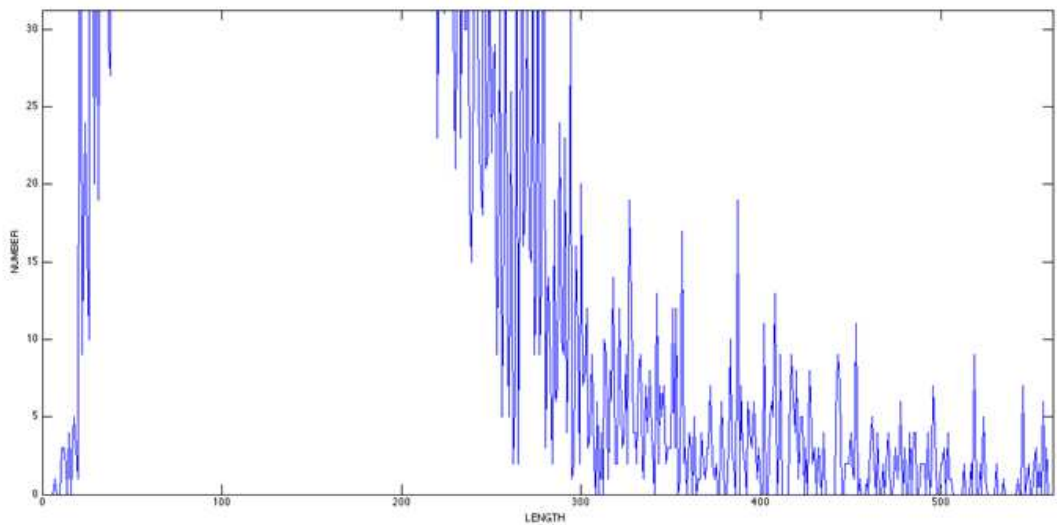}
\caption{partial enlargement of figure \ref{fangfa1} }\label{fangfa2}
\end{figure}

In figure 4 and figure 5, $s(40-300)/S=0.9471$, which means that exons with length between $40$ and $300$ are made up $95\%$ of all the exons approximately. Therefore, we select sequences with length of $40$ because we only consider characters of exon start. This length can contain the information of bases in exon start as much as possible and also, this length is less than the length of a whole exon.

\subsection{Selection by Using SVM}
\subsubsection{Support Vector Machine (SVM)}
In machine learning, support vector machines are supervised learning models with associated learning algorithms that analyze data and recognize patterns, used for classification and regression analysis. Given a set of training examples, each marked as belonging to one of two categories, an SVM training algorithm builds a model that assigns new examples into one category or the other, making it a non-probabilistic binary linear classifier. An SVM model is a representation of the examples as points in space, mapped so that the examples of the separate categories are divided by a clear gap that is as wide as possible. New examples are then mapped into that same space and predicted to belong to a category based on which side of the gap they fall on.\\

The SVM solves the problem of local extrema that cannot be avoided in neural network. SVM also have many advantages such as preventing overfitting effectively, applying in large feature space and compressing given information or dataset.[2] \\

Also, the main benefits of the SVM are the following three aspects:\\
\begin{enumerate}
  \item The SVM can find optimal solution under existing information, especially when sample is limited.
  \item The training process is to solve the problem of quadratic form optimization. Theoretically, we can get the global optimal point.
  \item The SVM could obtain high-dimensional feature by nonlinear transformation (kernel function) and create a linear discriminant function in this feature space to achieve the nonlinear discrimination in original space.
\end{enumerate}

At the same time, the SVM solves the dimensional problem because the complication of algorithm is unrelated to the dimension of sample.[2]  Due to strict theoretical foundation, there are many successful application in bioinformatics such as the identification of splice sites, the identification of start codon and the differentiation of host and pathogen.[2]   The SVM performs better in identifying result than traditional machine learning methods.[3]

\begin{itemize}
  \item Mathematical principle \\
According to given dataset,

\[T=(x_1,y_1),(x_2,y_2),\cdots,(x_L,y_L)\in(X*Y)\]
\[    y_i=
\begin{cases}
1 & \text{if $x_i$ is in given group}\\
-1 & \text{if $x_i$ is in other group}
\end{cases}\]
\[x_i \in X=R^n, y_i\in Y, i=1,\cdots , L \]

In which, $X$ is called input space. Every single point $x_i$ in the input space has n characters. Then we find a real-valued function $g(x)$ on $R^n$. By using classification function $f(x)=sgn(g(x))$, the value of $y$ corresponding to $x$ can be got and that is saying, a classification problem.

  \item Linear SVM \\

  We consider training set $T$. If $\exists \omega\in R^n, b\in R$ and positive number $\epsilon$, s.t. for all $i$ where $y_i=1$ we have $(\omega* x_i `) +b<=-\epsilon$. Then we say that training set $T$ is separable and the classification problem is separable.
\end{itemize}

\subsubsection{Support Vector Machine (SVM)}
A sequence with $40$ bases maps to a $1*160$ vector. The mapping is A to $(0,0,0,1)$, G to $(0,0,1,0)$, C to $(0,1,0,0)$ and T to $(1,0,0,0)$. After training SVM in MATLAB, we label a sequence of $40$ bases in exon start as $+1$ and a sequence of $40$ bases in intron start, which indicates the exon ending, as $-1$, to selected candidate exon start for further research.\\

Table \ref{SCIFIES} shows selected candidate initial and final exon sections of $No.48$ DNA sequence in GENESCANtest65 by applying SVM. The numbers in this table represent the location of base. The length of candidate exon bases ranges from $40$ to $300$.

\begin{table}[!h]
\centering
\begin{tabular}{|c|c|}
  \hline
  exon start & exon ending \\
   \hline
  94 & 371 \\
   \hline
  121 & 371 \\
   \hline
  264 & 371 \\
   \hline
  276 & 371 \\
   \hline
  295 & 371 \\
   \hline
  574 & 844 \\
   \hline
  583 & 844 \\
   \hline
  624 & 844 \\
   \hline
  660 & 844 \\
   \hline
  670 & 844 \\
   \hline
  $\cdots$ & $\cdots$ \\
  \hline
\end{tabular}
\caption{selected candidate initial and final exon sections}\label{SCIFIES}
\end{table}

We selected a DNA sequence of GENESCANtest65 randomly. The figure 6 shows the predicted candidate exon sections by using our algorithm, plotted by the location of base in DNA sequence on the horizontal axis and frequency of the base occurring in candidate exon on the vertical. The highlighted sections represent true exons.

\begin{figure}[!h]
\centering
\includegraphics[width=1\linewidth]{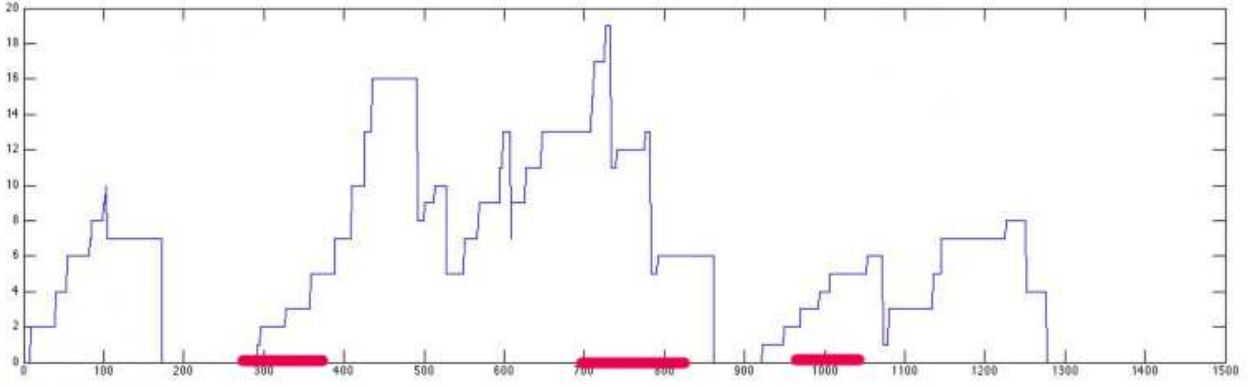}
\caption{predicted candidate exon sections}\label{SVM3}
\end{figure}

\subsection{DNA Numerical Value Mapping}
We mapped DNA sequence to numerical sequence by VOSS mapping. Let I={A, T, G, C}, then any DNA sequence of length N can be expressed as
\[S={S(n), \space \space \space S(n)\in I, n=0, 1, 2, \cdots, N-1}\]
To map symbol sequence to discrete numerical sequence, for any given $b\in I$,
Let
\[u_b[n]=\begin{cases}1\quad S(n)=b\\0\quad S(n)\neq b\end{cases}\quad n=0,1,2,\cdots,N-1\]
Take given sequence ATGCTTAG as an example. Its numerical value mapping shows in table \ref{num} .

\begin{table}[!h]
\centering
\begin{tabular}{|c|c|c|c|c|c|c|c|c|}

  \hline
  DNA sequence & A & T & G & C & T & T & A & G \\
  \hline
  $U_A$ & 1 & 0 & 0 & 0 & 0 & 0 & 1 & 0 \\
  \hline
  $U_T$ & 0 & 1 & 0 & 0 & 1 & 1 & 0 & 0 \\
  \hline
  $U_G$ & 0 & 0 & 1 & 0 & 0 & 0 & 0 & 1 \\
  \hline
  $U_C$ & 0 & 0 & 0 & 1 & 0 & 0 & 0 & 0 \\
  \hline
\end{tabular}
\caption{numerical value mapping of VOSS mapping}\label{num}
\end{table}

\subsection{Discrete Fourier Transform and Three-base Periodicity}
Combining base sequence selected by biological characteristics in the first step and AG sequence labeled as +1 and GT sequence labeled as -1 in SVM analysis, we obtain candidate exonic sections. After applying DFT, frequency analysis is adopted in numerical gene sequence. Using three-base periodicity, we can predict exon sequence.\\

To study the characteristics of DNA coding regions, DFT is adopted in indicator sequence.
\[U_b[k]=\sum_{n=0}^{N-1} u_b[n] e^{-i\frac{2\pi nk}{N}}, k=0,1,\cdots ,N-1\]
From this, we get four sequences of complex numbers of length $N$, $\{U_b[k]\}$,$b\in I$. Calculating the square of power spectrum of every sequence of complex numbers $\{U_b[k]\}$ and adding up them, we get the sequence of power spectrum, $\{P[k]\}$, of the whole DNA sequence $S$.
\[P[k]=\mid U_A[k]\mid ^2 +\mid U_T[k]\mid ^2 +\mid U_G[k]\mid ^2 +\mid U_C[k]\mid ^2\quad k=0,1,\cdots ,N-1\] \\

Most of existing prediction approaches based on digital signal processing techniques rely on the phenomenon that protein-coding regions have a prominent power spectrum peak at frequency $f = 1/ 3$ arising from the length of codons. But for introns there is no such phenomenon.\\

Noted the mean of power spectrum of DNA sequence S as $\bar{E}=\frac{\sum_{k=0}^{N-1}P[k]}{N}$. The ratio of the value of power spectrum at $N/3$ of DNA sequence to the mean of power spectrum of DNA sequence $S$ is Signal to Noise Ratio (SNR), which is $R=\frac{P[N/3]}{\bar{E}}$. When the SNR is more than an appropriate threshold value $R_0$ (such as $R_0=2$), this indicates coding sequence of DNA, exon. And introns do not have such property. When DNA sequence is relatively long, DFT takes lots of time for calculation. In fact, the frequency distribution of four nucleotides can give the value of power spectrum at $N/3$ of DNA sequence and the SNR directly. Let $x_b$,$y_b$,$z_b$ represent the frequency of nucleotides occurring at three locations of a codon, that is saying, $x_b$,$y_b$ and $z_b$ represent the occurrence frequency of nucleotide $b$ on sequence$0$,$3$,$6$,$\cdots$ and $1$,$4$,$7$,$\cdots$ and $2$,$5$,$8$,$\cdots$ correspondingly. Then we get the value of power spectrum at $N/3$ of DNA sequence, that is

\begin{align}\nonumber
P[N/3]& =\sum_{b\in I}\mid U_b[\frac{N}{3}]\mid ^2=\sum_{b\in I}\mid \sum_{n=0}^{N-1}u_b[n]e^{-j\frac{2\pi n\frac{N}{3}}{N}}\mid ^2\\\nonumber
& =\sum_{b\in I}\mid \sum_{n=0}^{N-1}u_b[n]e^{-j\frac{2\pi}{3}n}\mid ^2=\sum_{b\in I}\mid x_b+y_be^{-j\frac{2\pi}{3}}+z_be^{j\frac{2\pi}{3}}\mid ^2\\\nonumber
& =\sum_{b\in I}(x_b^2+y_b^2+z_b^2-x_by_b-x_bz_b-y_bz_b)
\end{align}

Obviously, when $x_b$, $y_b$ and $z_b$ approach to equal, the value of power spectrum at $N/3$ also approaches to zero. So the power spectrum curve of exon sequence has a prominent peak at frequency $f = 1/ 3$. This also reflects the disequilibrium of four nucleotides in exon sections.

\section{Algorithm Testing}

Here we assume that the threshold value of SNR is $R_0$. When $R\geq R_0$, we say that it is an exon and when $R\leq R_0$, it is an intron. Then, we will compare our predicted result and real condition. To evaluate our predicted result, we introduce four parameters shown as below.\\
\\
TP (true positive): The number of nucleotides that are in coding regions and are predicted correctly.\\
\\
TN (true negative): The number of nucleotides that are in non-coding regions and are predicted correctly.\\
\\
FN (false negative): The number of nucleotides that are in coding regions and are predicted falsely.\\
\\
FP (false positive): The number of nucleotides that are in non-coding regions and are predicted falsely.\\
\\
And an example of applying these four parameters shows as figure 7.
\begin{figure}[!h]
\centering
\includegraphics[width=1\linewidth]{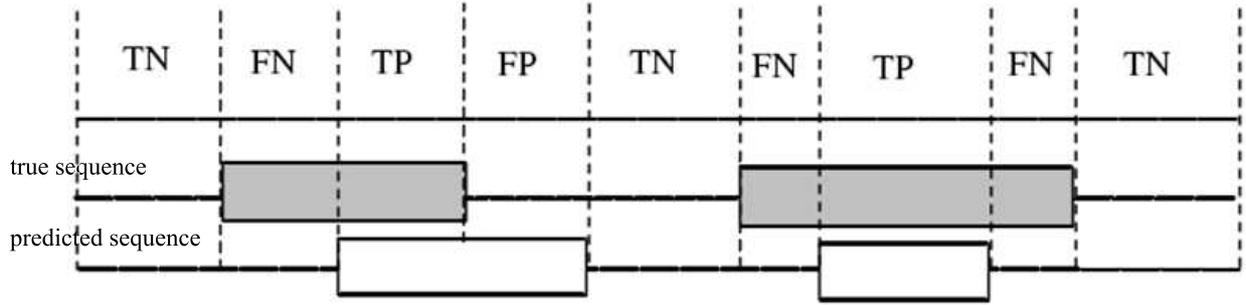}
\caption{example of applying four parameters}\label{SNSP}
\end{figure}

Based on TP, TN, FP and FN, we introduce evaluation index, sensitivity $S_n$ and specificity $S_p$.\\
\\
Sensitivity $S_n: S_n = \frac{TP}{(TP+FN)}$\\
\\
Specificity $S_p: S_p = \frac{TP}{(TP+FP)}$\\
\\
Coding regions and non-coding regions are shown by figure 8.[4]

\begin{figure}[!h]
\centering
\includegraphics[width=0.7\linewidth]{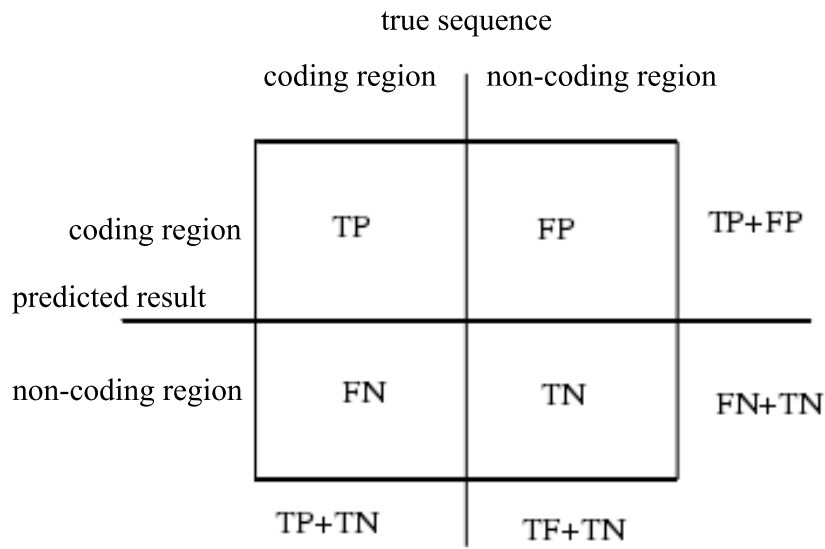}
\caption{coding regions and non-coding regions shown by TP,TN,FN,FP}\label{SNSP2}
\end{figure}

The sensitivity indicates true positive ratio of the algorithm and the specificity represents false positive ratio of the algorithm. Our objective is to let sensitivity and specificity to be maximum by giving a certain threshold value. To evaluate the effect of different threshold value, we plot the receiver operating characteristic (ROC) curve.See figure 9 .

\begin{figure}[!h]
\centering
\includegraphics[width=0.7\linewidth]{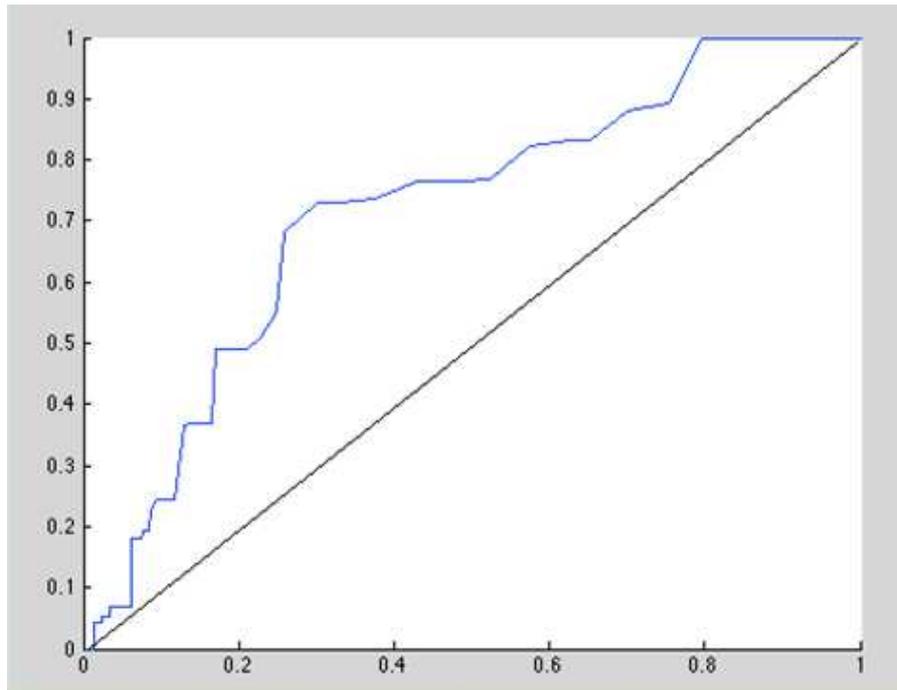}
\caption{ROC curve}\label{ROC}
\end{figure}

In this graph, $1-S_p$ is plotted on the horizontal axis and $S_n$ on the vertical. The area under ROC curve (AUC) is $0.7239$.

\section{Summary and Prospect}

The gene prediction method in this article combines biological characteristics of DNA, the SVM theory and three-base periodicity to avoid some potential drawbacks of any single theory. This method is supposed to enhance the accuracy of predicting protein coding regions, which is important to interpret the genetic characteristics of organisms. We select candidate exons according to universally applicable biological characteristics, which gain a foundation of biology and conform to the rule of scientific research. In the next step, to differentiate candidate exon start and intron start, we trained SVM for further selection. The SVM has a wide scope of application and higher efficiency compared with traditional algorithms such as Fourier Transform. We can obtain a relatively accurate classification outcome from the SVM. The DFT is adopted on selected candidate exon after above steps and the three-base periodicity is also employed to predict exons accurately. In this step, we use well-understood method of signal processing for a higher predicting accuracy.\\

Due to huge amount of data, bioinformatics has been an important research field in modern biology. Through twenty years of painstaking efforts, there are significant achievements in gene prediction. Gene prediction has developed from identifying protein coding regions in bacterial genome to prediction detailed structure of vertebrate gene. The essential issue of computer aided gene prediction is to predict the accurate location of gene in genome sequence when certain gene sequence has been given. By far, there are dozens of algorithms for predicting protein coding regions and ten algorithms or so and relevant software are provided for free on the Internet.[5][6]\newline
\newline
\newline
\begin{center}
  \LARGE{A}\large{CKNOWLEDGMENT}
\end{center}
The authors would like to thank Prof. Yihui Luan and Dr. Guangchen Liu for helpful discussions and pointers in early versions of this work, and a careful reading of the manuscript and also, thank the anonymous reviewers for their helpful comments and suggestions.

\bibliographystyle{plain}

\end{document}